\documentclass[10pt,a4paper, twocolumn]{article}
\usepackage[top=0.85in,footskip=0.75in,marginparwidth=2in]{geometry}

\usepackage[utf8]{inputenc}

\usepackage{cite}

\usepackage{nameref,hyperref}

\usepackage[right]{lineno}

\usepackage{microtype}
\DisableLigatures[f]{encoding = *, family = * }

\usepackage{changepage}

\usepackage[aboveskip=1pt,labelfont=bf,labelsep=period,singlelinecheck=off]{caption}

\makeatletter
\renewcommand{\@biblabel}[1]{\quad#1.}
\makeatother

\usepackage{lastpage,fancyhdr,graphicx}
\usepackage{epstopdf}
\fancyheadoffset[L]{2.25in}
\fancyfootoffset[L]{2.25in}

\usepackage{color}

\definecolor{Gray}{gray}{.25}

\usepackage{sidecap}

\usepackage{wrapfig}
\usepackage[pscoord]{eso-pic}
\usepackage[fulladjust]{marginnote}
\usepackage{amsmath} 
\usepackage{amssymb}
\usepackage{algorithm}
\usepackage{algorithmic} 
\usepackage{subfig} 

\usepackage{amsfonts}

\usepackage{authblk}

\usepackage{booktabs} 
\usepackage{colortbl}
\usepackage{multirow}

\reversemarginpar
\usepackage{atbegshi}
\AtBeginDocument{%
	\rfoot{\copyright This manuscript version is made available under the CC-BY-NC-ND 4.0 license  \\ https://creativecommons.org/licenses/by-nc-nd/4.0/}%
	\AtBeginShipoutNext{%
		\rfoot{}%
	}%
}

\begin{document}
	\twocolumn[\begin{@twocolumnfalse}
		\begin{flushleft}
			{\Large
				\textbf\newline{The Soccer Game, bit by bit:  An information-theoretic analysis}
			}
			\newline
			\\
			Luis Ramada Pereira\textsuperscript{1*},
			Rui J. Lopes\textsuperscript{2, 3},
			Jorge Louçã\textsuperscript{1},
			Duarte Araújo\textsuperscript{4},
			João Ramos\textsuperscript{5,4},
			\\
			\bigskip
			\bf{1} ISTAR Instituto Universitário de Lisboa (ISCTE - IUL) Lisbon, Portugal
			\\
			\bf{2} IT-IUL Instituto de Telecomunicações, Lisbon, Portugal
			\\
			\bf{3} Instituto Universitário de Lisboa, ISCTE-IUL, Lisbon, Portugal
			\\
			\bf{4} CIPER, Faculdade de Motricidade Humana, Universidade de Lisboa, Lisboa, Portugal
			\\
			\bf{5} Faculdade de Ciências da Saúde e Desporto, Universidade Europeia, Lisboa, Portugal
			\\
			\bigskip
			*ramada.pereira@iscte-iul.pt
			
		\end{flushleft}

		\section*{Abstract}
			In this article, we present an original method to measure the rate of 
			positional change observed during a soccer match based on the relative spatial distribution of players on the pitch. This is justified as players use their relative position as a key tactical tool to contribute to their team's objectives. A temporal network representation of the game was used, where nodes are players discretely clustered by physical proximity into disjoint clusters. This study, observational and descriptive in nature, was applied to a set of matches from a major European national football league, with players' coordinates sampled at 10Hz, resulting in $\approx$ 60,000 network samples per match. We took an information theoretic approach to measuring the distance between successive samples. Significant correlations were found between measurements and key match events that are empirically known to result in players jostling for position, such as when striving to get unmarked or to mark. These events increase the information distance between samples, while breaks in game play have the opposite effect. Having a measurement of dynamic, structural change in soccer is an original contribution that can complement full match statistical analysis. Hierarchical decomposition of the measurements is possible at multiple levels, building an overall multi-layer map that provides insights into the game dynamics, from the individual player, to the clusters of interacting players, up to the teams and their matches. This comprehensive view of the players' interacting behavior can be useful for training, tactics and strategy development. 
\\
\\
\end{@twocolumnfalse}
]

\section{Introduction}
\label{introduction}
\noindent
	Complex systems, with time evolving interactions among its elements, abound in the social, biological and physical domains.  In many of these systems, 
	elements are clustered in groups that also undergo changes with time. A temporal, clustered network can be an appropriate representation of such a system. 
	
	In this article we apply this representation to the sport of soccer. Soccer, as many other competitive team sports, can be seen as a socio-biological complex system. The domain dynamics of agent behavior in these sport modalities are neither fully random nor fully designed  \cite{Ramos2020}. This contributes decisively to their complexity. Agents cooperate and compete in clusters towards shared and conflicting goals. These clusters are frequently functionally bounded, such as in the group interactions of forward and defense players, or goal keepers and strikers. It is common knowledge that self-organization in complex systems emerges from constrained local action, so this representation appears, in principle, justified. A comprehensive discussion of the application of complex systems theory to football can be found in \cite{Salmon2020}.     
	
	The soccer match is represented in this article as a succession of network observations where clusters are subsets of players, including the two football goal frames, resulting in a network with a maximum of 24 nodes concurrently active, plus substitutes \cite{Ramos2017}.  
	
	While studying a soccer match as an evolving clustered network, we start from the proposition that players' spatial distribution is the determining variable for clustering. Not in relation to the pitch boundary, but in relation to their teammates and adversaries.  This is reasonable if we consider that being marked or unmarked, supported or unsupported, has a major impact on the opportunities for action that a player enjoys. The research question that this study aims to answer is whether the changes these clusters suffer as the game evolves, promise a more faithful indication of game dynamics when compared to traditional measurements such as the ratio of successful passes, speed, distance covered, and others.        
	
	Intuitively, we could think that an optimal assignment of players to clusters would require a physical distance measure, predicating link weights by player relative distance. However, there are complicating factors to the usage of such a precise measurement, as the importance of inter-player distance is not independent of game play \cite{Ramos2018}. It varies with pitch location, ball position, game rules, environmentals (such as playing surfaces or weather), or the relation between time and distance in dynamic game settings. All these contribute to the actual player's instantaneous grasp of his performance environment and perception of opportunity for action \cite{Araujo2016}. 
	
	This was the basis that lead us to cluster players and goals into homogeneous and disjoint groups connected by a single link \cite{Ramos2017}, using the formalism of hypergraphs \cite{Berge1973}. A hypergraph is characterized by having multiple nodes connected by a single link, called a hyperedge, in contrast with a traditional graph where links have a maximum of two endpoints. A set of nodes that share a link is called a simplex.  In the particular context of the present article, simplices are sets or clusters, and the collection of simplices observed in a single sample, a clustering. 
	
	We use the term ``clustering'' to mean the set of disjoint non-empty subsets of nodes observed in the network at a given point in time. Some authors call it a ``partition''. These terms represent similar constructs, clustering being semantically associated with  an emerging, bottom-up aggregation of nodes, while partition conveys the idea of a  top-down driven process. In soccer there is not a single entity controlling group formation \cite{Ribeiro2019b}, at least not directly and in real time, so the former seems more appropriate. 
	
	In the restricted context of this article, simplices and clusters are synonyms, both referring to the same construct: a group of players in articulated interaction and proximity. An example of the clustering process is illustrated in figure \ref{fig:7} in the appendix.

	It could be argued that discretization and assignment of nodes to a pairwise disjoint family of sets, would lead to a distorted representation of events on the pitch. After all, players move freely in an Euclidean space and in continuous real time, while in the proposed representation time is discrete and players move on a lattice, understood not as a grid that spans the pitch but as the configuration space of all possible set arrangements \cite{Conway1999, Johnson2010}. Frequent observation, however, mitigates these effects. For example, peripheral players in a simplex will more easily transfer to a different simplex and, if frequently observed, any simplex changes will be quickly captured. Due to the high frequency characteristic of the network (10Hz), errors will smooth out as player simplices form and dissolve, establishing a bridge between the continuous domain of game play and the time sliced network representation employed \cite{Johnson2016}. 
	
	This discretization carries with it a significant advantage. We are no longer in a continuous domain, and the toolkit of information theory \cite{Cover2006} becomes available to us. In a discrete domain, information can be quantified for complexity, such as in the Kolmogorov complexity or the Shannon entropy \cite{Kolmogorov1968, C.E.Shannonvol.27pp.JulyOctober1948, Grunwald2008}. Similarly, two pieces of information can be compared for distance. We can determine how far apart or how close they are by the number of units of information that are needed to find one given the other. In this article the pieces of information are the individual clustering samples of the soccer match. We measure their distance using the \textit{Variation of Information}, an entropic based metric introduced by Marina Meilǎ in 2003 \cite{Meila2003}, to compare clusterings. A detailed description and reasons for selection can be found in section \ref{TFU}. It's on this intersection of network science and information entropy that this article is rooted.
	
	In the reminder of this document, we discuss related work in section \ref{rw}. Theoretic underpinnings, including major theories, concepts, key variables and the way they inform observations, correlation of $\mathit{VI}$ and playing dynamics and procedures used are in section \ref{methods}, which is followed by a section \ref{results} describing  our findings. We discuss these results in section \ref{discussion} and we conclude with directions for future research in section \ref{conclusion}.

	\section{Related work}
	\label{rw}
	Using networks and entropic measures to study the soccer game is not new. In this section we refer to prior studies that have explored these techniques and explain how they differ from this article's approach. This is not a comprehensive review or description of networks or entropy and their use. The reader is referred to \cite{Ribeiro2017} for a summary of the implications and merits of applying network science to team sports performance analysis, and to \cite{Ribeiro2021}, where a description of the extensive variants of entropy, some of which have been used in team sports analysis,  can be found. 
	
	In comprehensive reviews of the literature, such as those found in \cite{Lord2020,Sarmento2018} where authors analyze performance and general research trends in soccer and other team invasion sports, networks are a popular topic. In \cite{Sarmento2018}, a review fully dedicated to soccer, 11.7\% of articles reviewed use networks and network metrics as an analysis tool, and in \cite{Lord2020}, a review of the literature on performance analysis of team invasion sports, 10.8\% of the reviewed articles focusing on soccer make use of network analysis. All of these articles  use exclusively networks built out of dyadic interactions between players on ball passing and crossing, sometimes incorporating spatio-temporal metrics \cite{Cotta2013,Gama2014,Clemente2016a}. Usually a weighted digraph is built per team, sometimes broken down to individual attacking play \cite{Korte2019}, and statistics such as clustering coefficient, network density, centrality or degree distribution are used to explain patterns of play or performance. Spatial analysis is accomplished dividing the pitch into diverse zones, either longitudinally or on both axis, and assigning arcs (i.e. directed links) connecting the passes' origin and target zones. Specific attacking plays, such as those ending up in a scored goal have also been analyzed using these techniques \cite{Mclean2018}.
	
	Entropy has been previously used to study soccer dynamics, but much less frequently than network science. As an example, in the reviews referenced above, there are only two explicit references to articles dealing with soccer and entropy. 
	
	In \cite{Vilar2013} authors clustered players by their location in seven pitch sectors, dynamically bounded by the 20 outfield players. Similarly to our approach, this clustering is performed every 0.1s. They then computed the difference in the number of players from each team in each of the sectors and measured the Shannon entropy of its frequency for the whole match, resulting in an uncertainty measurement of local dominance. This was used to identify correlates of performance and patterns of intra and inter team coordination, understood as the level of sector numerical dominance that results from player interactions. Although the sectors, and thus the clusters, are dynamically defined, there is a level of inflexibility by fixing the number of clusters of players per observation. The clustering method also does not avoid assigning players in closer interaction to separate clusters. In contrast with the entropic measure used in this article, it prevents fine grained temporal analysis, as it is frequency based.
	
	In \cite{Lopes2019} Shannon entropy (among other information theoretic measures) is used to study multiple national leagues using rounds as time units, with home and away goals as variables. The authors found the emergence of similar entropy patterns across seasons and across leagues. 
	
	In \cite{Couceiro2014} authors quantified space coverage variability of players, by discretizing the pitch area into 1 $m^2$ cells and using the frequency distribution of players over the cell map to compute its Shannon entropy. As expected they found that midfielders exhibit a higher entropy than other players. According to the authors, this result is more ``assertive'' than a typical heat map. Approximate entropy, a time series analysis technique that can reveal the predictability of patterns, was also used in this article, to analyze the distance covered by a defender. It was possible to categorize the respective time series (at 1s interval) as a chaotic system, somewhere in between periodic and random.      
	
	Approximate entropy, was also used to analyze spatial statistics, such as occupied areas, dispersion or team center of gravity in \cite{Duarte2013a}. No clustering of players was performed and time analysis was limited to 15 min segments. The same technique  was also used in \cite{Sampaio2012}, with different spatial and dynamical properties, to study the effect of tactical training in a group of student footballers playing small sided matches.  
	
	In \cite{Neuman2018, Martinez2020, Ribeiro2020} we find examples of studies that use networks in conjunction with entropic measures in match analysis. In \cite{Ribeiro2020}, authors used the same network formalism and representation as used herein, and sample entropy to measure the synchronization between players, their simplices and teams, from a time series of observed cluster phases. They observed different axial synchronization of player-simplex phases, on two small sided games setups with different conditions of goals', number, sizes and location (4 mini-goals without goalkeepers versus 2 larger goals with goalkeepers). 
	
	\cite{Neuman2018, Martinez2020} are both  based on pass networks. In \cite{Neuman2018} authors used the Tsallis entropy, a measure that generalizes the traditional Boltzmann-Gibbs/Shannon entropy to non-extensive systems (that is, systems where sub-states are not mutually independent), to study its  correlation with team performance and season results. The analysis is performed at match level, and the authors found that, under certain parametrization, the Tsallis entropy of a team is inversely correlated with team performance. The opposite result is observed when considering the difference of team entropies per match. In \cite{Martinez2020} authors performed a spatial entropy analysis of pass origins at match level, and a temporal analysis of network parameters with high correlation with the number of passes, such as the longitudinal coordinate of center of mass of the pass network or the network clustering coefficient, using permutation entropy on a time varying series built with a moving window of 50 passes. 
	
	The network design approach we took for this article diverges substantially from a passing network. It is self-evident that only a player in possession of the ball can score, which is a strong argument in favor of using passing networks for performance analysis. However, as pointed out in \cite{Grund2012}, relevant interactions in a soccer game are not limited to passes. Intuitively, the opportunity for a successful pass is perceived by the player carrying the ball, as a function of multiple variables, in which the dynamic position of some of his teammates and adversaries play a major role. The same can be said for the opposing team while trying to intercept or clear a pass. As mentioned in \cite{Hewitt2016} ``Players must be able to pass with precision while others create space around themselves to receive the pass from their teammate''. It is dynamics like this that we try to capture by using the formalism previously introduced. In the specific case of passes, the temporal changes in clusterings are precursors for a passing opportunity or interception. In non formal language, we can say that in a passing network we can find what happens, while in a polyadic network of player's interactions, we can explain why it happens! 
	
	There are other differences in the proposed approach that circumvent some of the challenges of passing networks. Relations in passing networks are inherently dyadic, although, as mentioned, a player passing decisions are inherently polyadic. Passing networks are usually a single team view, where the influence of the opposing team is usually absent. Interceptions and clearances are ignored, although they may have a decisive impact on the game. The use of signed networks could address some of these difficulties, but introduce theoretical challenges, as many of the metrics of simple networks do not extend to signed networks, which is probably the reason that, to our knowledge, they have not been used for this purpose. And, finally, compared to positioning actions, passes are relatively rare, leading to a low temporal resolution when gathering statistics. In \cite{Yamamoto2011}, authors propose a minimum window size of 5 min, to collect passing data. The reader is referred to \cite{Buldu2018} for a thorough discussion of the challenges of using passing networks.
	
	The representational formalism used in this article was introduced in \cite{Ramos2017, Ramos2018}. In those articles, every match observation was partitioned into clusters of players in proximal spatial interaction, and several variables were extracted from this representation. Here, we extend this prior work to reveal the changes these clusters experience across time, and explore their meaning by using an information entropic metric.
	
	In summary, the major original contributions introduced in this article and detailed ahead, are:
	\begin{itemize}
		\item Using dynamic polyadic relations between players, more faithfully representing the player decision making process
		\item Measurements of cluster breakup and emergence that encompass home and away teams and their dynamics
		\item Structural change measurements that can be decomposed at multiple levels
		\item Change measurements without a fixed frame of reference, avoiding some of the pitfalls of traditional measurements.     
	\end{itemize}

	\section{Methods}
	\label{methods}  
	In this section we cover the theories, concepts, constructs, key variables, and the way they inform the observations in section \ref{TFU}, and the procedures used to represent and analyze the captured data from the sample set of matches in section \ref{Proc}.     
	\subsection{Theoretical Framework and Underpinnings}
	\label{TFU}
	Every observation of a match is a clustering of nodes, representing players and goals.
	Formally, a clustering is: 
	\begin{multline}
		C = \{c_1, \cdots, c_k\}: \\
		(c_i \cap c_j =  \varnothing  \;\; \forall \;( 1 \leq i, j \leq k\;
		\land \; i \neq j)) \land \; \cup _{i=1}^{\,k} \, c_i = V
		\label{eq:1} 
	\end{multline}
	where $c$ are the disjoint subsets, $k$ the number of subsets, and $V$ the set of all nodes.
	
	There are several methods to measure the inter-distance between clusterings, with varying properties, 
	such as the Rand Index \cite{Rand1971}, Adjusted Rand Index \cite{Hubert1985}, the Normalized Mutual Information \cite{Danon2005}, the Van Dongen-Measure \cite{Dongen2000} and others. A thorough discussion of the major methods can be found in \cite{Vinh2010,Wagner2007, Meila2007}.    
	We selected the Variation of Information $(\mathit{VI})$ \cite{Meila2007}, also known as Shared Information Distance, to measure the information distance between samples and thus evaluate the change a clustered network experiences as a function of time. 
	The choice of $\mathit{VI}$ is justified as it is a true metric, respecting the triangle inequality, meaning that no indirect path is shorter than a direct one. This is important in analyzing the rate of change at multiple scales, avoiding the unreasonable possibility of having a greater rate of change for a given time interval, when sampling the network at a lower rate. $\mathit{VI}$ also increases when fragmentation and merges occur in larger clusters, which intuitively relates to playing dynamics, given the rise in degrees of freedom experienced in larger groups of interacting players. Fundamentally, although in this article we consider $\mathit{VI}$ as a proxy for game dynamics, $\mathit{VI}$ itself is not a quantification of informational meaning or semantics, but simply, a quantification of informational variation, or as Shannon puts it “semantic aspects of communication are irrelevant to the engineering problem” \cite{C.E.Shannonvol.27pp.JulyOctober1948}. 
	
	In simple terms, $\mathit{VI}$, measures the amount of information required to obtain one clustering (observation) from another. 
	If no changes in the clusters are observed, then there is no variation of information. As clusterings shift from one another, $\mathit{VI}$ increases.  
	This is easy to visualize when considering the so-called confusion matrix \cite{Stehman1997} between clusterings at successive observations. This matrix describes the node spread, where each element represents the number of nodes moving from one cluster to another. If clusters are unchanged and keep their node affiliation, the confusion matrix will be a monomial matrix, $\mathit{VI}=0$ and we know exactly where each node ends up. But as the number of non-zero entries in the confusion matrix increases and their distribution tends to uniform, the uncertainty about each node destination also increases. Consider as an example a cluster that splits in half versus another that sheds a single node. There is a higher uncertainty about each node final destination in the former than in the latter. $\mathit{VI}$ measures this uncertainty.  A practical illustration of how to compute  $\mathit{VI}$  can be seen in tables \ref{tab:table1} and \ref{tab:table2} in the appendix.  
	
	Formally,  $\mathit{VI}$ is a function that takes two clusterings as parameters and returns the information distance between the clusterings. $\mathit{VI}$ is computed as:
	\begin{equation}
		\label{eq:3} 
		VI(X;Y) = - \sum_{i=1}^k\sum_{j=1}^lr_{ij}[\log_2(\frac{r_{ij}}{p_i})+\log_2(\frac{r_{ij}}{q_j})]
	\end{equation}
	where $X=\{x_1,\cdots, x_k\}$ and  $Y=\{y_1,\cdots, y_l\}$ are clusterings of a given set $S$, with $n=\vert S \vert$, $k=\vert X \vert$, $l=\vert Y \vert$, $r_{ij}=\frac{\vert x_i \cap y_j \vert}{n}$, $p_i=\frac{\vert x_i \vert}{n}$ and $q_j=\frac{\vert y_i \vert}{n}$. From this equation it is easy to see that when the clusters in $X$ and $Y$ are the same, the result is zero, as $r_{ij}=p_i=q_j$. This result expresses the fact that no information is gained or lost when going from one clustering to the other.
	For empty intersections of pairwise clusters, $r_{ij}=0$, and although $\log(0)$ is not defined, applying l'Hopital rule we get a null contribution from these intersections to the overall $\mathit{VI}$. In summary, only pairwise non-disjoint, non-identical clusters contribute to the information distance. This contribution led us to introduce an additional construct, the simplex transition. Simplex transitions can be statistically analyzed, and their frequency and contribution to overall  $\mathit{VI}$, can provide insights into structural change and dynamics of the match.           
	
	$\mathit{VI}$ works as a distance metric for clusterings of the same set of nodes. In the model used to represent the soccer match, the set of nodes remains constant, except on substitutions and send-offs. However, the number of observations affected by these events are so low, that we have ignored their contribution in the model. 
	
	Using base 2 logarithms, $\mathit{VI}$ is measured in bits (or shannons) and describes the balance of information needed to determine one clustering from another.   $\mathit{VI}$ is algorithmically simple (it can be computed in $\mathcal{O}(n + k l)$)) and, as mentioned before, it is a true metric \cite{Kraskov2005}, respecting positivity, symmetry, and the triangle inequality.
	
	Using the previous notation, for every individual player $p_{ij} \in \{x_i \cap y_j \}$ his contribution to the overall $\mathit{VI}$ is computed as:
	\begin{equation}
		\label{eq:4}
		\mathit{VI}^{p_{ij}} = -r_{ij}\frac{[\log_2(\frac{r_{ij}}{p_i})+\log_2(\frac{r_{ij}}{q_j})]}{\vert x_i \cap y_j\vert}
	\end{equation}
	which takes the contribution of pairwise clusters $x_i, y_j$ to the overall $\mathit{VI}$, and divides it in equal parts among all players $\in x_i \cap y_j$. Note that, in the particular case of the network that we built, all nodes/players are present in all observations and are members of one and only one cluster in any one observation. Equation \ref{eq:4} registers the contributions of players involved in their clusters when these change. The only exception is the case of a send-off or substitution, in which case the player no longer contributes to the dynamics of the match.     
	
	The $\mathit{VI}$ of two clusterings ($X, Y$) of $S$ can only be zero if $\forall s \in S \mid s \in X \leftrightarrow s \in Y$. If this condition is not met then $\min(\mathit{VI}) \geq \frac{2}{n^*}$ \cite{Meila2007}, where $n^* = \max(k, l)$ still using the same notation. In the soccer match representation proposed in this article the number of nodes  is fixed at 24 (barring any red cards), and thus, $n^*=12$ and $\min(\mathit{VI})=\frac{1}{12}$ every time there are any clustering changes. $\mathit{VI}$ depends on the level of fragmentation on the pitch across observations, which intuitively reflects the situation of players jostling for position, but cannot exceed $\log_2(n)$ \cite{Meila2007}. These extreme values of $\mathit{VI}$ are, however, just boundaries that limit minima and maxima given any set of clusterings. In the present case, we have a  minimum of 2 nodes per cluster, which implies a maximum of 12 clusters, resulting in $\max(\mathit{VI})=\log_2(12)=3.585$, which is attained when a clustering with a single cluster splits into 12 clusters with two nodes each, or vice-versa. In practice, the maximum VI registered is substantially lower with typical observed values of $\max{\mathit{VI}} \approx 1.2$, corresponding to the maximum distance between clusterings with $0.1s$ separation. 
	
	
	\subsection{Procedures}
	\label{Proc}
	
	The proposed framework was applied to the analysis of a set of 9 soccer matches from the 2010-11 season of the English Premier League. Based on an information stream collected from realtime pitch-located raw video feed, each match is modeled as a high-resolution (10Hz) temporal hypernetwork with simplices as clusters of players  and goals parsed by proximity. Each network is made up of up to 30 nodes (28 players and 2 football goals) of which only a maximum of 24 are present on the pitch at any given moment (11 players from each team and 2 goals). The inclusion of goals is justified when considering that the purpose of the polyadic formalism that we use is to capture the multiple factors that may affect a player's decision making process, and proximity to goals is certainly an important one.  The number of simplices is variable, dependent only on the observed map of players and goals. The method used for clustering guarantees that a node and its closest node belong to the same simplex, or, in other words, it guarantees that no node is closer to a node belonging to a different simplex than to its closest node in the same simplex.  This implies that the smallest simplex has a minimum of 2 nodes, i.e., there are no isolated nodes. Although there maybe occasions where a player is side-lined, this will be an exception, as the expectation at the top-level of sports performance is that every single player have an active role in-play, in relation to their teammates and their opponents. Although the football goals are obviously fixed on the pitch, there is no fixed frame of reference for the clustering process. The algorithm used for clustering is non parametric and is explained in \cite{Ramos2017}. 
	
	On average, considering a match, including extra time, we observed and measured the network $\approx$ 60,000 times. Each of these 60,000 samples is a clustering of the network. 
	
	The output of the method is a time series of  $\mathit{VI}$ measurements, that can be hierarchically decomposed into separate measurements for teams, players, and simplex transitions.

\begin{figure*}[!h] 
		\centering
		\includegraphics[scale=0.66, trim= 4 4 4 4]{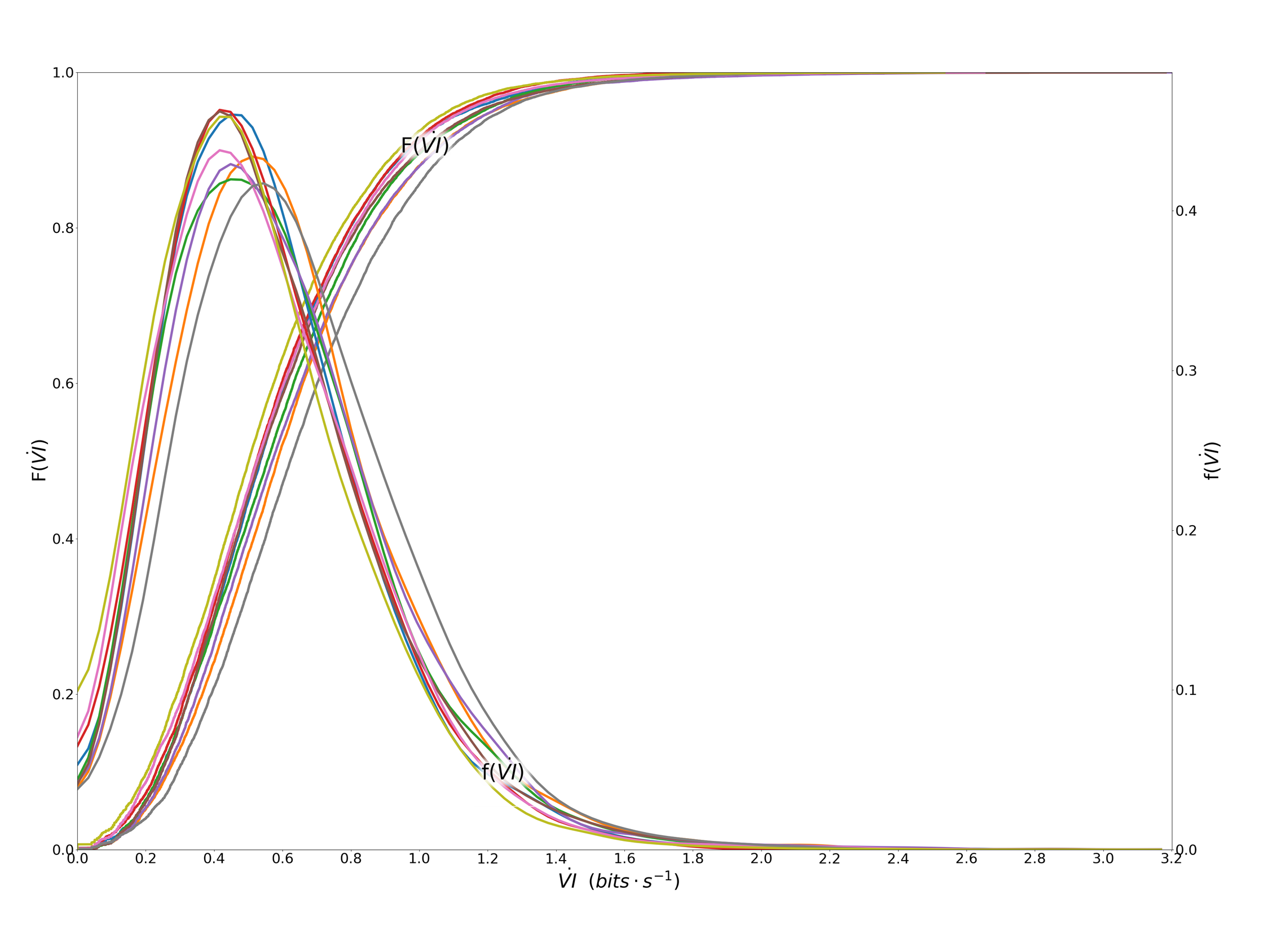}
		\caption{Probability Density Function ($f(\dot{\mathit{VI}})$) and Cumulative Distribution Function ($F(\dot{\mathit{VI}})$) for all nine matches measured on a 4s moving average window. Games color coded. There is a consistency of patterns that likely mirrors energy expenditure and management throughout the game \cite{Osgnach2010}.}
		\label{fig:0}
	\end{figure*}

	At 10Hz, a significant amount of sparsity, i.e. a large amount of transitions without clustering changes, is observed. This posits the question of the ideal sampling rate \cite{Moura2013}, given the dynamics of a soccer game, the capturing technology and the clustering methodology. The observed sparsity lead us to adopt a set of measures in the findings section ahead, to enhance analysis and observability. These included:
	\begin{itemize}
		\item the usage of differentials and measuring change in bps, denoted as $\mathit{\dot{VI}}$;
		\item the use of moving averages for visualization and compatibility with the rate of change and play of a soccer match. Results shown use 4s sample windows, except when noted; 
		\item and, finally, we made use of cubic Hermite splines \cite{Neuman1978} to envelope $\mathit{\dot{VI}}$ maxima. Results use an inter pivot distance that dynamically varies up to a maximum of 80s depending on the position of the observed value in the probability density function of $\mathit{\dot{VI}}$ (figure \ref{fig:0}).        
	\end{itemize}   

	\section{Findings}
	\label{results}
	In this section we present the key findings resulting from our analysis. 
	\subsection{Clusterings reappear much more frequently than expected by chance} 
	Given that the space of all clusterings is substantial, corresponding to a lattice of over $4.4\times 10^{17}$ points (Bell number $B_{24}$), the amount of unique clusterings we can observe is just a small fraction of this space, gated by the total of samples collected (average $58283$, $\sigma=1336$). Assuming a random distribution, the probability of observing the same clustering, that is the same sets of simplices, is for all purposes nil when considering the space size ($1.46\times10^{-12}$). Obviously the real distribution is not random and is heavily condition by its prior state. But, when excluding consecutive observations, a significant level of clustering re-appearances still emerges (average $6.4\%$, $\sigma=0.5\%$), which, intuitively, can be interpreted as the influence of strategic design over match playing patterns \cite{Ramos2020}. 
	\subsection{Different time series, similar statistics}
	Having analyzed nine soccer matches of the 2010-11 season of the English premier league at 10Hz, on a 40 sample moving average window (4s), we found that the average $\mathit{\dot{VI}}$ and the standard deviation for the whole match is consistent across matches, with a total average of $0.597$ bps, $\sigma=0.0369$. 
	
	Considering that a typical player spends on average over half of his time standing or walking and only sprints ($ > 8.3 ms^{-1}$) 1.4\% of the time \cite{Ferro2014}, 10Hz is a sampling frequency that often generates no clustering changes in consecutive samples. In fact, in almost 80\% of the network observations clusterings do not change.    
	The standard deviation per match has an average $\mathit{\dot{VI}}$ of 1.30 bps, with a maximum of 1.37 and a minimum of 1.25 bps across all nine matches. A full report for all matches can be found in table \ref{tab:table3}.
	
	The dispersion of $\mathit{\dot{VI}}$ as measured by the coefficient of variation of all match observations averages $218\%$, reflection of the high activity level of the soccer game.   
	
	We found no correlation between the time ordered sets of $\mathit{VI}$ observations between the matches we have analysed. 
	When comparing different matches, we found consistent $\mathit{\dot{VI}}$ averages, with a coefficient of variation of the averages of $\approx 5\%$. 
	
	The probability density function of a match $\mathit{\dot{VI}}$ measurements is highly consistent across matches as seen in figure \ref{fig:0}. Matches exhibit similar probabilities of finding given levels of dynamics and we did not find matches where  $\mathit{\dot{VI}}$ is consistently high or consistently low. An explanation is player's regulation of exertion  during the match to manage fatigue,  particularly at the high intensity professional matches are played \cite{Sarmento2018}. All matches come from the official English premier league games, usually played at a similar competitive level, so these  results are not surprising, if $\mathit{\dot{VI}}$ does  accurately reflecting game dynamics.

	\subsection{ Time decreasing trend of $\mathit{VI}$} 
	
	In 8 out of the 9 matches we examined, we observed a lower  $\mathit{VI}$ when comparing the second half to the first half. Neuromuscular, biochemical and perceptual changes leading to increased physical and mental fatigue as a match progresses has been extensively documented \cite{Silva2018}. More specifically, indicators such as total distance with the ball, high intensity running with the ball, among other typical indicators of performance have been shown to measure lower on the 2nd half of a match \cite{Rampinini2009}. Adjusted tactics, resulting from increased acquaintance with competitor behavior, may be a further compounding cause.  
	
	A reduction in physical match performance (high speed running and sprinting) has also been reported when comparing the first 15 minutes of the first and second half \cite{Weston2011}. In line with this report, in our sample we observed a lower $\mathit{VI}$ in all matches, under the same conditions.    
	
	However, it is important to note that in our sample the same team plays in every match. A larger sample of matches, from a wider population, may offer more consistency to this pattern, although these results already suggests a strong correlation between the proposed metric and game intensity, deserving further study. The observed values of $\mathit{VI}$ can be seen in table \ref{tab:table_fatigue}.
	
	\begin{table}[htbp]
		\centering
		\caption{Comparison of $\dot{\mathit{VI}}$ between 1st and 2nd half of 9 matches, and between first 15 minutes of each half. In only one match do we observe values (shaded red) contrary to reported trends of intensity indicators}
		\begin{tabular}{|c|cccc|}
			\midrule
			Match & \multicolumn{1}{p{3.6em}}{\textcolor[rgb]{ .247,  .247,  .463}{1st Half}} & \multicolumn{1}{p{3.6em}}{\textcolor[rgb]{ .247,  .247,  .463}{2nd Half}} & \multicolumn{1}{p{3.6em}}{\textcolor[rgb]{ .247,  .247,  .463}{15 min \newline{}1st Half}} & \multicolumn{1}{p{3.6em}|}{\textcolor[rgb]{ .247,  .247,  .463}{15 min 2nd Half}} \\
			\midrule
			1    & \textcolor[rgb]{ .247,  .247,  .463}{0.555} & \textcolor[rgb]{ .247,  .247,  .463}{0.533} & \textcolor[rgb]{ .247,  .247,  .463}{0.566} & \textcolor[rgb]{ .247,  .247,  .463}{0.533} \\
			2    & \textcolor[rgb]{ .247,  .247,  .463}{0.611} & \textcolor[rgb]{ .247,  .247,  .463}{0.571} & \textcolor[rgb]{ .247,  .247,  .463}{0.601} & \textcolor[rgb]{ .247,  .247,  .463}{0.532} \\
			3    & \textcolor[rgb]{ .247,  .247,  .463}{0.634} & \textcolor[rgb]{ .247,  .247,  .463}{0.628} & \textcolor[rgb]{ .247,  .247,  .463}{0.703} & \textcolor[rgb]{ .247,  .247,  .463}{0.623} \\
			4    & \textcolor[rgb]{ .247,  .247,  .463}{0.679} & \textcolor[rgb]{ .247,  .247,  .463}{0.650} & \textcolor[rgb]{ .247,  .247,  .463}{0.703} & \textcolor[rgb]{ .247,  .247,  .463}{0.693} \\
			5    & \textcolor[rgb]{ .247,  .247,  .463}{0.614} & \cellcolor[rgb]{ 1,  .8,  .8}0.630 & \textcolor[rgb]{ .247,  .247,  .463}{0.630} & \textcolor[rgb]{ .247,  .247,  .463}{0.624} \\
			6    & \textcolor[rgb]{ .247,  .247,  .463}{0.590} & \textcolor[rgb]{ .247,  .247,  .463}{0.556} & \textcolor[rgb]{ .247,  .247,  .463}{0.698} & \textcolor[rgb]{ .247,  .247,  .463}{0.584} \\
			7    & \textcolor[rgb]{ .247,  .247,  .463}{0.599} & \textcolor[rgb]{ .247,  .247,  .463}{0.539} & \textcolor[rgb]{ .247,  .247,  .463}{0.617} & \textcolor[rgb]{ .247,  .247,  .463}{0.518} \\
			8    & \textcolor[rgb]{ .247,  .247,  .463}{0.639} & \textcolor[rgb]{ .247,  .247,  .463}{0.559} & \textcolor[rgb]{ .247,  .247,  .463}{0.639} & \textcolor[rgb]{ .247,  .247,  .463}{0.546} \\
			9    & \textcolor[rgb]{ .247,  .247,  .463}{0.603} & \textcolor[rgb]{ .247,  .247,  .463}{0.558} & \textcolor[rgb]{ .247,  .247,  .463}{0.550} & \textcolor[rgb]{ .247,  .247,  .463}{0.547} \\
			\bottomrule

		\end{tabular}%
		\label{tab:table_fatigue}%
	\end{table}%
	
	\subsection{Notational event data correlates with $\mathit{VI}$}
	To validate the hypothesis that $\mathit{VI}$ is a measure of game dynamics, we searched for correlations between known moments of intensive player repositioning and surges in the information distance.
	Corners, being overwhelmingly defended one-to-one \cite{Pulling2013}, result in quick player displacement and occur frequently in a match (mean $10.3$, $\sigma=2.5$, which matches previously reported numbers \cite{Casal2015}). This justifies, in our view, the selection of corners for hypothesis validation. 
	
	We collected timed tags for corners from match commentary. These events are time tagged down to the minute of play. To address the different resolutions scales of  commentaries and clustering samples, we computed, per match,  the mean $\dot{\mathit{VI}}$ for every minute of play, and compared its median with the mean for the minutes when corners were taken.  Out of 93 corners, 86 had a higher $\mathit{VI}$ than the median. The probability of this occurring on random chance is $1.33\times10^{-16}$.   
	
	%

	\begin{figure*}[!ht]
		\centering
		\includegraphics[scale=0.27]{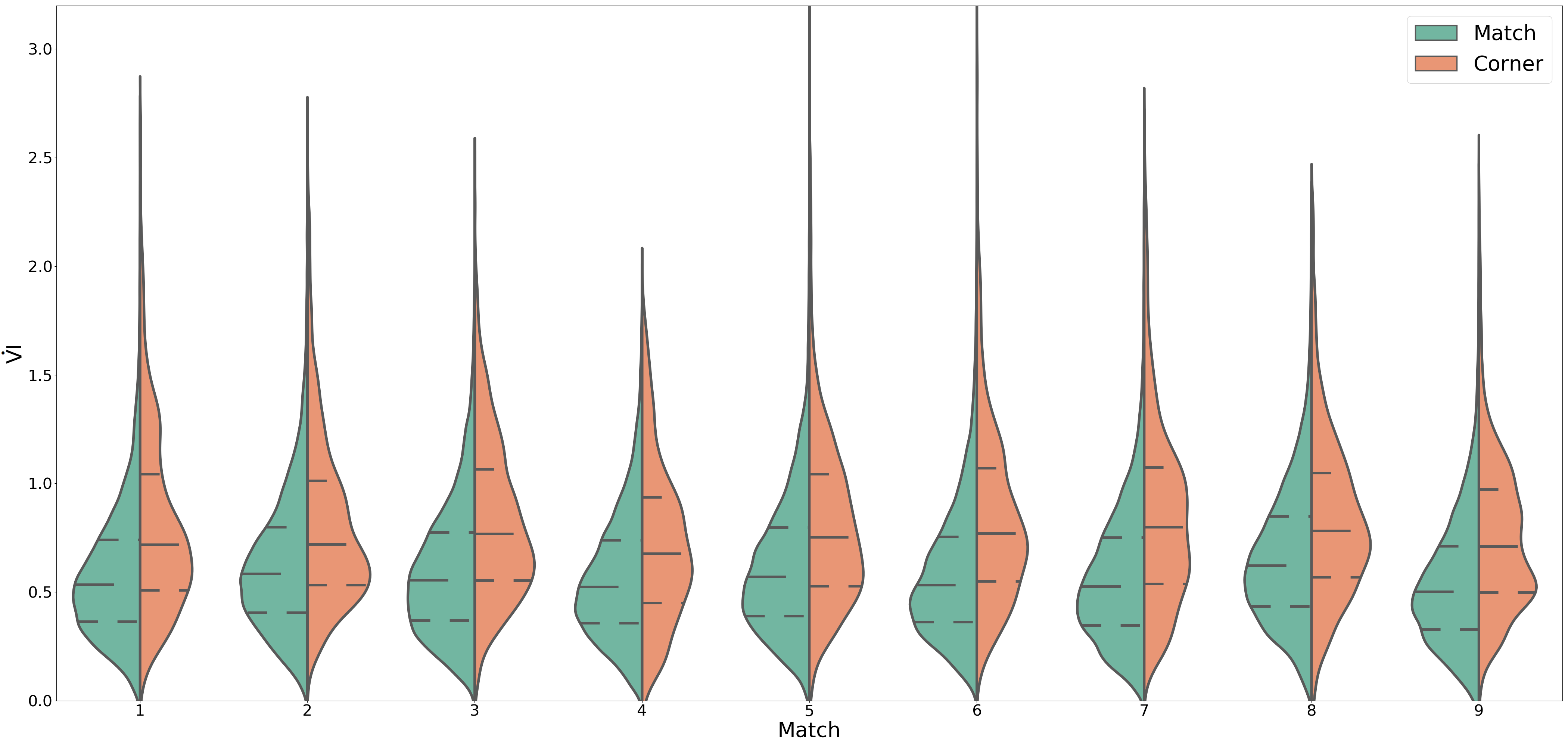}
		\caption{Distribution of $\dot{\mathit{VI}}$ for a complete match, compared with observations during the minutes when corners are taken, using a Gaussian kernel density estimate. There is a notable skew towards higher $\dot{\mathit{VI}}$ observations during corners for all nine matches.}
		\label{fig:10}	
	\end{figure*}

	We also inspected the $\dot{\mathit{VI}}$ distribution for the whole match and compared to the same distribution for all corners' minutes. As can be seen in figure \ref{fig:10}, all matches show a $\dot{\mathit{VI}}$ distribution that is skewed higher when comparing with match averages.

\begin{figure*}[!hbtp] 
		
		\centering
		\subfloat[][Match 1, 0-0]{
			\includegraphics[scale=0.47]{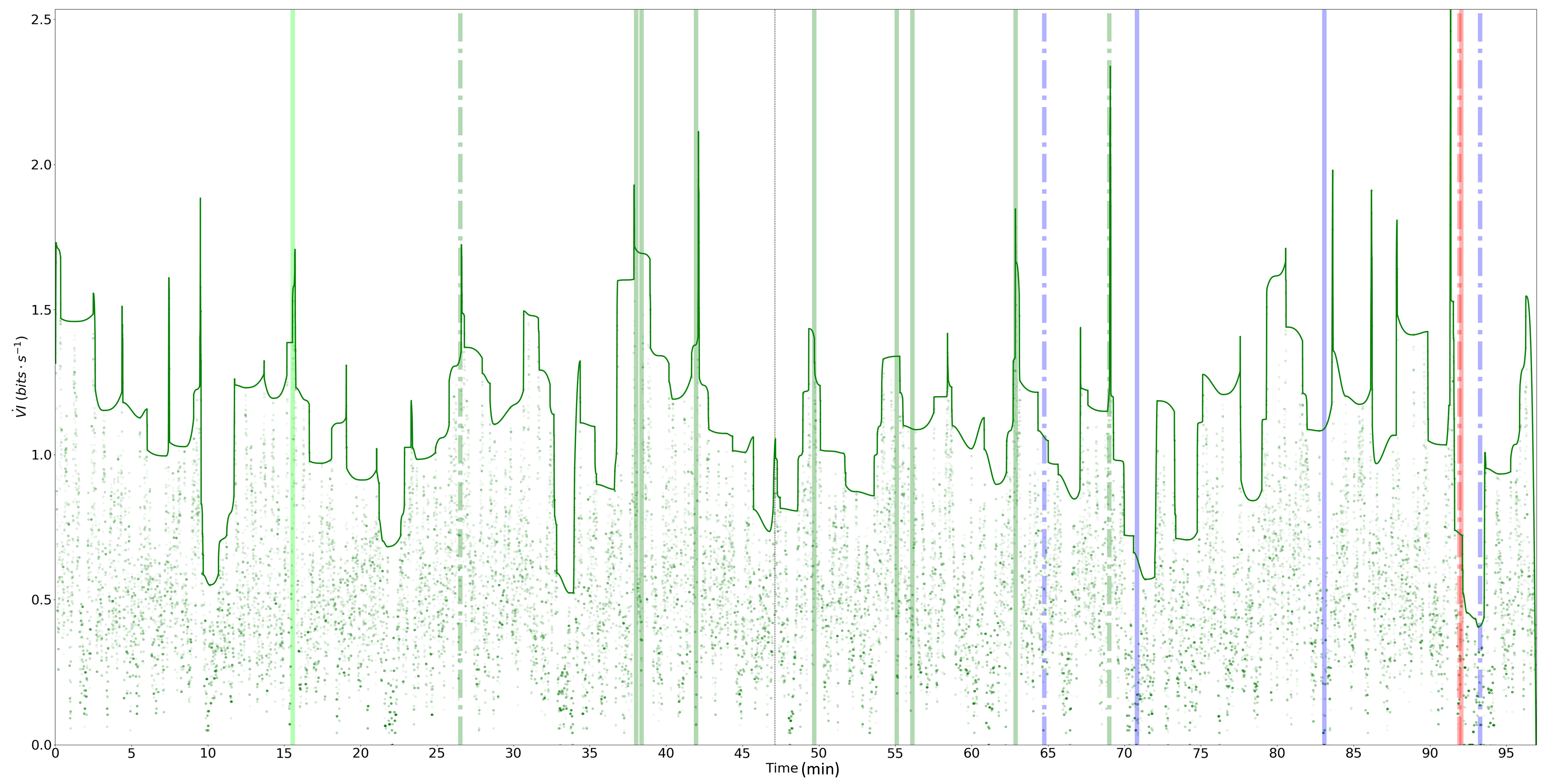}
			\label{sfig:31}}
		\\
		\subfloat[][Match 2, 2-1]{
			\includegraphics[scale=0.47]{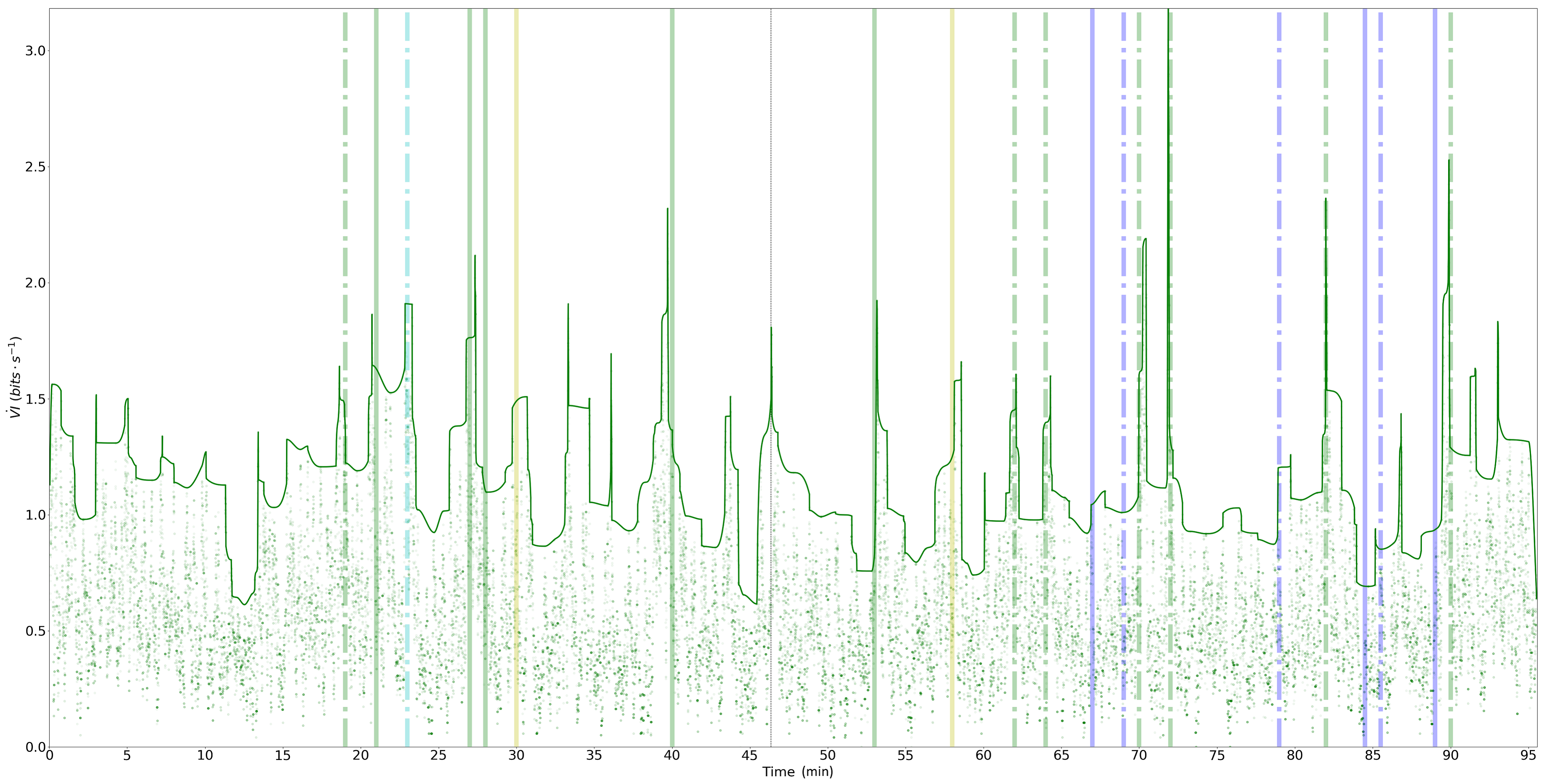}
			\label{sfig:32}}
		\\			
		\subfloat{
			\includegraphics[scale=0.64]{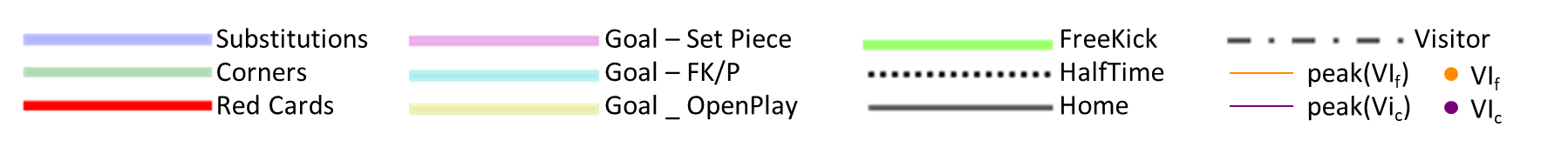}}
		\caption{Plots for two matches where green points are observations of $\dot{\mathit{VI}}$ at each sample transition, and the colored line the respective peak envelope. $\dot{\mathit{VI}}$ seems to be heavily correlated with match events, such as corners, where a high level of player repositioning is expected, and player substitutions, usually associated with a trough in $\dot{\mathit{VI}}$. It is also visible at minute 92 in \ref{sfig:31} that the match virtually "stopped" during the send-off of two players from opposing teams.}
		\label{fig:3}
	\end{figure*}

	These results provide compelling evidence that corners do indeed result in a marked increase of $\mathit{VI}$.  
	$\mathit{VI}$, as used in this study, is clearly a proxy for game dynamics, understood as a rapid pace of
	inter-players relative displacement, i.e. without a fixed frame of reference. This is notably obvious during set pieces. Corners and free kicks invariably generate a spike in $\mathit{VI}$. Conversely, other events, like substitutions or send-offs, generate pauses that are captured by a drop in $\mathit{VI}$. 
	Examples can be seen in figure \ref{sfig:31} and \ref{sfig:32}, where $\dot{\mathit{VI}}$ is plotted for a whole match, with vertical bars indicating the type and time of events.

	\subsection{Most simplex transitions occur only once} 
	We also introduce the concept of a simplex transition, a tuple of simplices $(c_i^{t}, c_j^{t+\delta})$ such that $(c_i^t \cap c_j^{t+\delta} \neq \varnothing) \land (c_i^t \neq c_j^{t+\delta})$, that, at successive observations, involves always the same players. 
	
	Most simplex transitions occur only once during a match. However there are some that occur with higher frequency (up to 50 times a match). These are usually symmetrical. They may be candidates for further analysis given their relative importance. In figure \ref{fig:9} the top contributing transitions of one match are represented, indicating their relative $\mathit{\dot{VI}}$ weight, the nodes involved, and when during the match they occurred.  
	
	\subsection{Player's $\mathit{VI}$ contribution for simplex transitions is related to his role } 
\begin{figure*}[!hbtp] 
		\centering
		\includegraphics[scale=0.10]{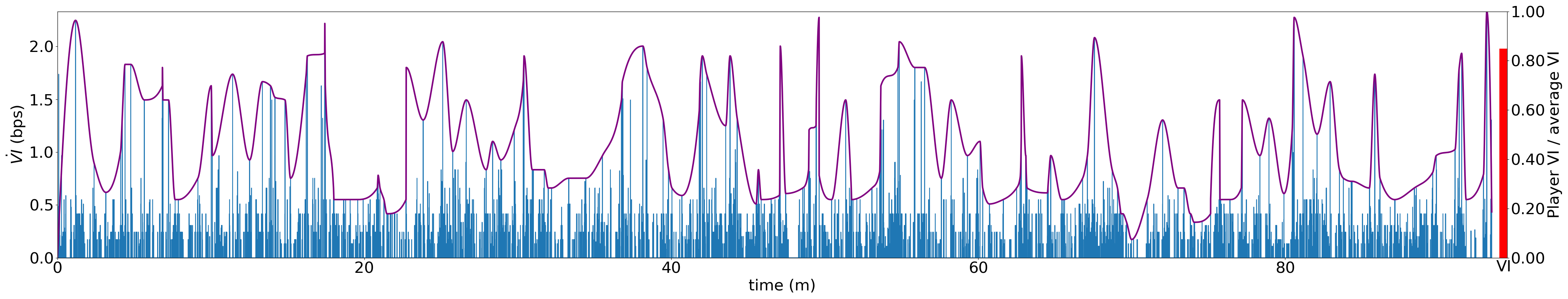}
		\caption{$\mathit{\dot{VI}}$ for a single player, in a single match, with maxima envelope. His total $\mathit{\dot{VI}}$ is compared against the match average for the whole match on the red bar on right hand side of this plot. In this case, a center forward player is represented, showing a lower than average $\mathit{\dot{VI}}$, which may be expected, because a forward is typically less active than the other players during his team defensive sub-phases of the match.} 
		\label{fig:8}
	\end{figure*}

	To analyse a player contribution to the overall $\mathit{VI}$, we apply equation \ref{eq:4}. His individual $\mathit{\dot{VI}}$, can be compared to the average $\mathit{\dot{VI}}$ per player. This may be useful to assess his activity during the match (figure \ref{fig:8}). Beyond the trivial low $\mathit{VI}$ observed for the goalkeepers, we observed anecdotal differences between forward, midfielders and defenders consistent with literature reports \cite{DiSalvo2007}.   	
	
	\begin{figure*}[!hbtp] 
		\centering
		\includegraphics[scale=0.300]{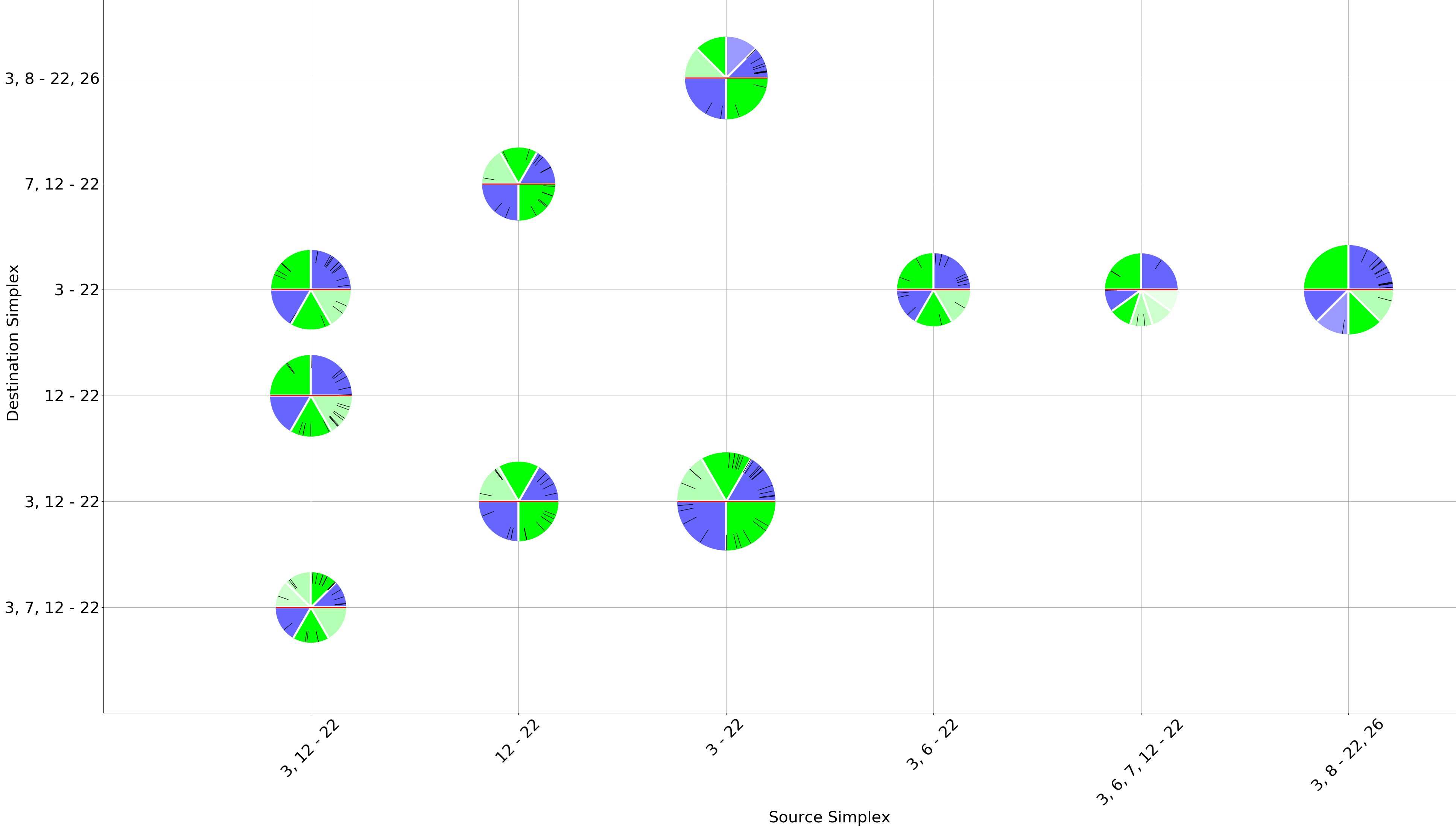}
		\caption{This chart shows the top ten simplex transitions player 22 of match 1 (figure \ref{sfig:31}) was involved in, as well as their formation. His contribution to the match $\mathit{VI}$  resulting from participating in these simplex transitions, is proportionally encoded in the area of the circle: larger circle signifies higher contributions. Each formation is coded in color and shade, with green and blue representing, respectively, home and visitor players, and the number of shades the number of participating players in the simplex. Each tick signals a transition and the match moment when it occurred, with a full match taking a full circle. The lower and upper semicircles describe, respectively, the formation of the prior (source) and immediately subsequent (destination) simplices, where the player was involved. Finally, simplices are identified by the participating players' numbers, with home players first, followed by visitors.
			Player 22 is a visiting forward, and as seen in the picture, is frequently observed alone (the single shade of blue in the semi circles) in a simplex with opposing back player(s), a typical pattern.  Transition from formation $3-22$ to $3,12-22$, when home player 12 joins the simplex, has the highest accumulated $\mathit{VI}$ contribution from player 22. It occurs throughout the match but with an emphasis in the first half of the first 45 min. Player 22 is supported by a teammate in only two transitions out of the 10 represented. } 
		\label{fig:4}
	\end{figure*}
	
\begin{figure*}[!ht] 
		\centering
		\includegraphics[scale=0.300]{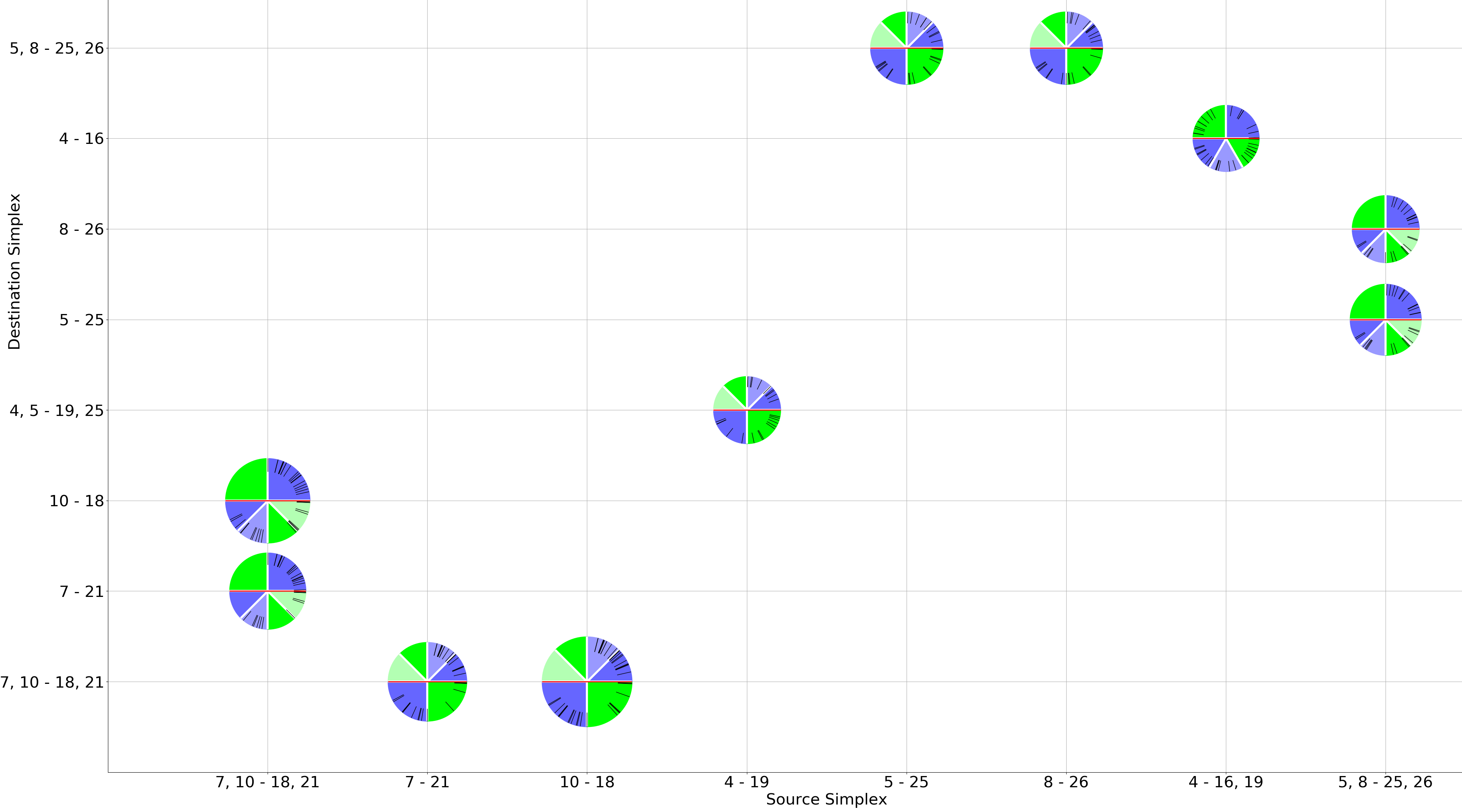}
		\caption{This chart uses the same symbolic elements as figure \ref{fig:4} but operates at a different level. Each circle represents the overall contribution to the match $\mathit{VI}$ of a whole transition and not just the player's contribution. Here we represent a match top ten transitions. The encoded information in this and in figure \ref{fig:4} can be useful to study and train high frequency transitions that contribute significantly to playing dynamics.} 
		\label{fig:9}
	\end{figure*}

	We visualize the type of transition, color coded to denote the number of home and visiting players involved. Each simplex transition plot is scaled by overall $\mathit{VI}$ contribution for that set of transitions, and details when those transitions occurred (see figure \ref{fig:9}). 
	
	\section{Discussion} 
	\label{discussion}
	
	A player's performance is dependent on how he perceives and responds to environmental cues that emerge from game play \cite{Travassos2014}. These cues, such as relative positioning of teammates, of adversaries and of the goal, condition the affordances the player has for action, while his own actions change this landscape. This feedback loop generates a complex system that researchers have striven to describe and understand.   By using a temporal network of ``relationships'' to represent the game, we endeavor to uncover insights otherwise missed. The correlations we observed  with moments of well known dynamics, endorse our approach. Corners are a prime example, but other events such as free strikes close to the penalty area or interruptions, correlate as well. Other reported observations, such as the impact of fatigue, of the halftime interval or of player role in intensity indicators, were also consistently detected in the 9 matches we analyzed. Although, the study could benefit from a larger sample, the evidence gathered, as shown, is certainly promising. 
	
	The proposed way of measuring the soccer game enables a multi-layer decomposition of its dynamics from macro level (a full match) to meso (clusters of players, transitions and teams), to micro (individual players). This enriches the information that can be extracted,  helpful to evaluate the dynamics generated by individual players, but also by sets of interacting players, which can uncover which players' structures are more prevalent, how they change and how they impact the overall match dynamics. It is also possible to inspect which simplex transitions a player is involved in, and split his contribution among simplex transitions as shown in figure \ref{fig:4}. An aggregation of all simplex transition charts provides a full view of a complete match. 
	
	As we stated in the introduction, this study is essentially descriptive in nature. This does not mean that the measurements we presented cannot be used for performance analysis. We should however be aware of what the authors in \cite{David2015} stated: ``A greater number of sprints by individuals in a team, amount of ball-related activities, or distance covered had no association with the probability of winning	matches''. It is true that our method avoids ``unproductive'' intensity, such as sprints that do not change relative positioning, or, other technical actions that do not increase the agency possibilities a player enjoys. However, given the impact of fatigue, instead of using directly $\mathit{VI}$ as introduced, considering its rate to player and team's work, could intuitively produce a more faithful predictor of performance.

	\section{Final Remarks and Future Research}
	\label{conclusion}
	
	The presented results endorse the status of  $\mathit{\dot{VI}}$ as a measure for game dynamics. The fact that it captures with accuracy and precision well known moments of players jostling for position, supports this interpretation. 
	
	With error free and detailed metadata, a more accurate analysis would be possible, especially with concurrent notation hard to capture automatically. The present work is based on prior data, captured and clustered independently, that abstract the reality of a soccer match. Based on the promise shown by the use of information theory and networks as analysis tools, the proposed methods could be valuable to evaluate different approaches to data capture, such as sampling rates, as well as different clustering methods and game representations, such as overlapping, distance weighted networks, non-inertial frames of reference that can accommodate ancillary factors, centroid based clustering, among many others. Extensions to multi-layer networks, where ball action can be integrated, could provide an additional level of insights.         
	
	All this is left for future research.           
	
	\clearpage
	
	
	\section*{Declarations}
	
	\subsection*{Acknowledgements}
	N/A
	
	\subsection*{Funding}
	This project was partly supported by Fundação para a
	Ciência e Tecnologia through project UID/Multi/04466/2019.
	R. J. Lopes was partly supported by the Fundação para
	a Ciência e Tecnologia, under Grant UIDB/50008/2020 to Instituto
	de Telecomunicações.
	D. Araújo was partly funded by Fundação para a Ciência e Tecnologia, grant number UIDB/00447/2020 attributed to CIPER – Centro Interdisciplinar para o Estudo da Performance Humana (unit 447).

	\subsection*{Abbreviations}
	N/A
	
	\subsection*{Availability of data and materials}
	N/A
	
	\subsection*{Ethics approval and consent to participate}
	N/A
	
	\subsection*{Competing interests}
	The authors declare that they have no competing interests.
	
	\subsection*{Consent for publication}
	N/A
	
	\subsection*{Authors' contributions}
	N/A
	
	\subsection*{Authors' information}
	N/A
	

	\clearpage
\onecolumn{\ifx\undefined\BySame
	\newcommand{\BySame}{\leavevmode\rule[.5ex]{3em}{.5pt}\ }
	\fi
	\ifx\undefined\textsc
	\newcommand{\textsc}[1]{{\sc #1}}
	\newcommand{\emph}[1]{{\em #1\/}}
	\let\tmpsmall\small
	\renewcommand{\small}{\tmpsmall\sc}
	\fi

	\clearpage
	\onecolumn
	\section*{Appendix}
	\label{appendix}
	To illustrate how $\dot{\mathit{VI}}$ is computed, consider the two moments in a fictional match represented in figure \ref{fig:7}. 
	The corresponding confusion matrix, which describes the transition of nodes between simplices when going from moment t to t+0.9s during the match, is given in table \ref{tab:table1}. 
	Null matrix elements, as well as unchanged simplices (simplices 1, 2 and 9), do not contribute to informational distance. The contribution of the others is computed according to equation \ref{eq:4}. The result is shown in table \ref{tab:table2}, where the contribution from each simplex transition can be seen.
	
	The end result is $\mathit{VI}=0.785615$ or, given that we are measuring a 0.9s interval, $\dot{\mathit{VI}}=\frac{0.785615}{0.9}=0.872905$ bps.
	
	\begin{table}[!h]
		\begin{tabular}{|c|c|c|c|c|c|c|c|c|c|}
			\toprule
			\multicolumn{1}{|l|}{\textbf{Simplex}} & \textbf{1} & \textbf{2} & \textbf{10} & \textbf{11} & \textbf{12} & \textbf{13} & \textbf{14} & \textbf{15} & \textbf{9} \\
			\midrule
			\rowcolor[rgb]{ .851,  .882,  .949} 1     & 2     & 0     & 0     & 0     & 0     & 0     & 0     & 0     & 0 \\
			\midrule
			2     & 0     & 2     & 0     & 0     & 0     & 0     & 0     & 0     & 0 \\
			\midrule
			\rowcolor[rgb]{ .851,  .882,  .949} 3     & 0     & 0     & 3     & 0     & 0     & 0     & 0     & 0     & 0 \\
			\midrule
			4     & 0     & 0     & 2     & 0     & 0     & 0     & 0     & 0     & 0 \\
			\midrule
			\rowcolor[rgb]{ .851,  .882,  .949} 5     & 0     & 0     & 0     & 0     & 0     & 0     & 0     & 3     & 0 \\
			\midrule
			6     & 0     & 0     & 0     & 2     & 1     & 0     & 0     & 0     & 0 \\
			\midrule
			\rowcolor[rgb]{ .851,  .882,  .949} 7     & 0     & 0     & 0     & 0     & 1     & 2     & 0     & 0     & 0 \\
			\midrule
			8     & 0     & 0     & 0     & 0     & 0     & 0     & 3     & 1     & 0 \\
			\midrule
			\rowcolor[rgb]{ .851,  .882,  .949} 9     & 0     & 0     & 0     & 0     & 0     & 0     & 0     & 0     & 2 \\
			\bottomrule
		\end{tabular}%
		\caption{Confusion matrix going from t to t+0.9s}
		\label{tab:table1}%
	\end{table}%
	\begin{table*}[!h]
		\begin{tabular}{|c|c|c|c|c|c|c|c|c|c|}
			\toprule
			\multicolumn{1}{|l|}{\textbf{Simplex}} & \textbf{1} & \textbf{2} & \textbf{10} & \textbf{11} & \textbf{12} & \textbf{13} & \textbf{14} & \textbf{15} & \textbf{9} \\
			\midrule
			\rowcolor[rgb]{ .851,  .882,  .949} 1     & 0     & 0     & 0     & 0     & 0     & 0     & 0     & 0     & 0 \\
			\midrule
			2     & 0     & 0     & 0     & 0     & 0     & 0     & 0     & 0     & 0 \\
			\midrule
			\rowcolor[rgb]{ .851,  .882,  .949} 3     & 0     & 0     & 0.092121 & 0     & 0     & 0     & 0     & 0     & 0 \\
			\midrule
			4     & 0     & 0     & 0.110161 & 0     & 0     & 0     & 0     & 0     & 0 \\
			\midrule
			\rowcolor[rgb]{ .851,  .882,  .949} 5     & 0     & 0     & 0     & 0     & 0     & 0     & 0     & 0.05188 & 0 \\
			\midrule
			6     & 0     & 0     & 0     & 0.048747 & 0.107707 & 0     & 0     & 0     & 0 \\
			\midrule
			\rowcolor[rgb]{ .851,  .882,  .949} 7     & 0     & 0     & 0     & 0     & 0.107707 & 0.048747 & 0     & 0     & 0 \\
			\midrule
			8     & 0     & 0     & 0     & 0     & 0     & 0     & 0.05188 & 0.166667 & 0 \\
			\midrule
			\rowcolor[rgb]{ .851,  .882,  .949} 9     & 0     & 0     & 0     & 0     & 0     & 0     & 0     & 0     & 0 \\
			\bottomrule
		\end{tabular}%
		\caption{Computing $\mathit{VI}$}
		\label{tab:table2}%
	\end{table*}%

	\begin{table*}[!hbtp]
		\begin{tabular}{|c|c|r|r|r|r|r|r|r|r|r|}
			\toprule
			\multicolumn{2}{|c|}{Match} & \multicolumn{1}{c|}{1} & \multicolumn{1}{c|}{2} & \multicolumn{1}{c|}{3} & \multicolumn{1}{c|}{4} & \multicolumn{1}{c|}{5} & \multicolumn{1}{c|}{6} & \multicolumn{1}{c|}{7} & \multicolumn{1}{c|}{8} & \multicolumn{1}{c|}{9} \\
			\midrule
			\multicolumn{2}{|c|}{Result} & \multicolumn{1}{c|}{0-0} & \multicolumn{1}{c|}{2-1} & \multicolumn{1}{c|}{2-2} & \multicolumn{1}{c|}{1-0} & \multicolumn{1}{c|}{3-0} & \multicolumn{1}{c|}{1-0} & \multicolumn{1}{c|}{0-1} & \multicolumn{1}{c|}{2-1} & \multicolumn{1}{c|}{1-0} \\
			\midrule
			\rowcolor[rgb]{ .851,  .882,  .949}  & Avg & 0.544 & 0.591 & 0.631 & 0.665 & 0.622 & 0.573 & 0.568 & 0.599 & 0.581 \\
			\rowcolor[rgb]{ .851,  .882,  .949} & $\sigma$ & 1.255 & 1.278 & 1.346 & 1.369 & 1.330 & 1.276 & 1.273 & 1.292 & 1.282 \\
			\rowcolor[rgb]{ .851,  .882,  .949}
			\multirow{-2}[2]{*}{$\dot{\mathit{VI}}_t$} & a & -4.6E-4 & -6.0E-4 & -2.9E-4 & -9.9E-4 & 1.4E-4 & -1.2E-3 & -8.7E-4 & -1.3E-3 & -4.7E-4 \\
			\midrule
			\multirow{3}[2]{*}{$\dot{\mathit{VI}}_h$} & Avg & 0.277 & 0.290 & 0.329 & 0.330 & 0.314 & 0.284 & 0.301 & 0.302 & 0.292 \\
			& $\sigma$ & 0.702 & 0.691 & 0.774 & 0.756 & 0.746 & 0.696 & 0.739 & 0.717 & 0.711 \\
			& a & -6.2E-5 & -4.2E-4 & 2.4E-4 & -3.6E-4 & -9.4E-5 & -6.2E-4 & 4.4E-4 & -6.3E-4 & -2.9E-4 \\
			\midrule
			\rowcolor[rgb]{ .851,  .882,  .949}  & Avg & 0.267 & 0.301 & 0.303 & 0.335 & 0.308 & 0.289 & 0.267 & 0.301 & 0.289 \\
			\rowcolor[rgb]{ .851,  .882,  .949} & $\sigma$ & 0.677 & 0.715 & 0.718 & 0.769 & 0.734 & 0.712 & 0.673 & 0.719 & 0.709 \\
			\rowcolor[rgb]{ .851,  .882,  .949} 
			\multirow{-2}[2]{*}{$\dot{\mathit{VI}}_v$} & a & -4.0E-4 & -1.8E-4 & -5.3E-4 & -6.2E-4 & 2.4E-4 & -6.2E-4 & -1.3E-3 & -6.6E-4 & -1.8E-4 \\
			\bottomrule
		\end{tabular}%
		\caption{Average (avg), standard deviation ($\sigma$), and linear regression slope (a) for  $\dot{\mathit{VI}}$ results (Total, Home and Visitor) for the nine matches used in this article}
		\label{tab:table3}%
	\end{table*}%
	\clearpage
	
	\begin{figure*}[!hbtp] 
		\centering
		\subfloat[][Clustering at time t]{
			\includegraphics[scale=0.45]{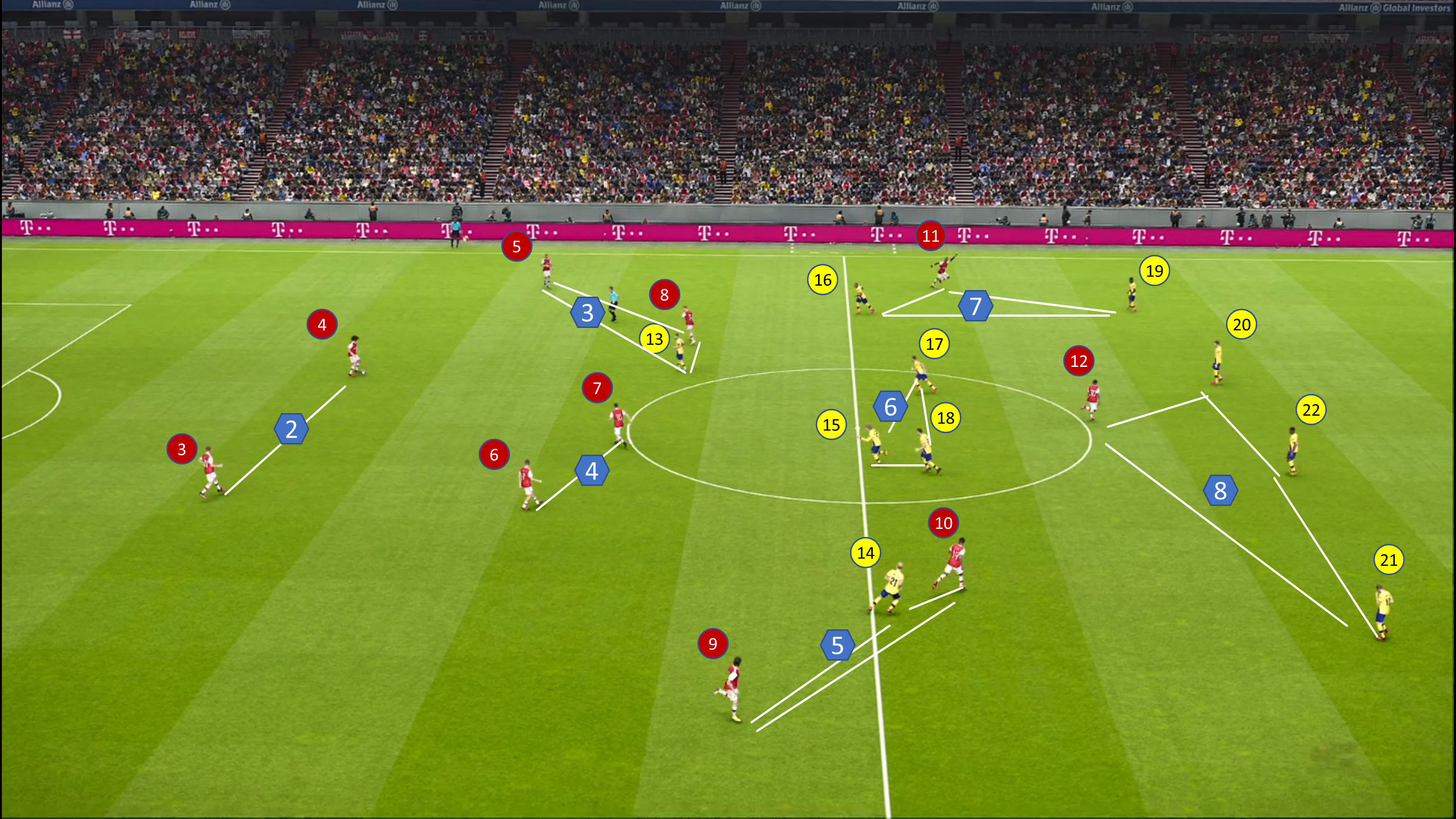}
			\label{sfig:71}}
		\\
		\subfloat[][Clustering at time t+0.9s]{
			\includegraphics[scale=0.45]{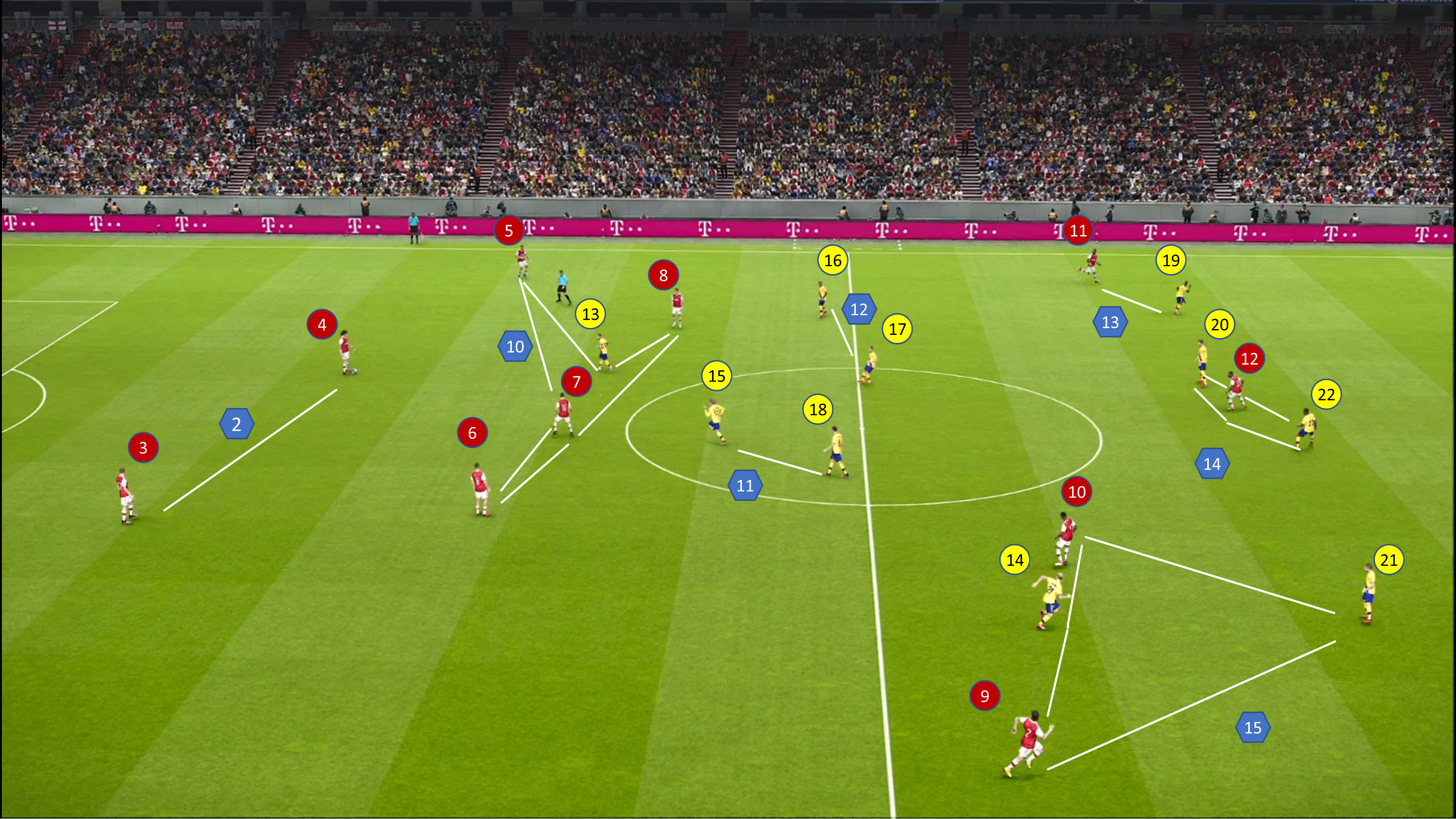}
			\label{sfig:72}}
		\\			
		\caption{Clustering for two moments of a fictional match separated by 900ms. Cluster 1 (goal and goalkeeper of the red team) and Cluster 9 (goal and goalkeeper of the yellow team), are not visible. The clustering process ensures that a node and its closest neighbor are nodes of the same simplex. Home players are numbered in red circles, visitors in yellow. Blue hexagons identify the simplices. White lines are only used to identify simplex membership. Formation for (a) is $\{2^4,3^4,4\}$ and for (b) is $\{2^6,3,4,5\}$, which correspond to the row and column sums of the matrix in table \ref{tab:table1}.}    
		\label{fig:7}
	\end{figure*}
	
\end{document}